\documentclass[aps]{revtex4}
\usepackage{epsfig, amssymb, latexsym, amsfonts, amsmath, amsthm}

\newcommand{\twz}{ {\text{tw2}} }

\def \d{{\mathrm{d}}}

\newcommand{\D}{\mathbf{d}}
\newcommand{\pd}{ \partial }

\newcommand {\kln}[1]{\left( #1 \right)}
\newcommand {\Kln}[1]{\bigl( #1 \bigr)}
\newcommand {\KLn}[1]{\Bigl( #1 \Bigr)}

\newcommand {\kle}[1]{\left[ #1 \right]}
\newcommand {\Kle}[1]{\bigl[ #1 \bigr]}

\newcommand {\kleo}[1]{\left[ #1 \right.}

\newcommand {\okle}[1]{\left. #1 \right]}

\newcommand {\im}{{\text{i}}}
\newcommand {\e}{{\text{e}}}

\begin{document}

\vspace{2cm}

\title{The twist-2 Compton operator  and \\
its hidden Wandzura-Wilczek and Callan-Gross structures\\
\phantom{relation}}

\author{ Bodo Geyer }
\affiliation{ Institute of Theoretical Physics,
Leipzig University, Augustusplatz~10, D-04109~Leipzig, Germany}

\author{ Dieter Robaschik }
\affiliation{BTU Cottbus,
Fakult\"at 1, Postfach 101344, D-03013~Cottbus, Germany }

\date{\today}

\begin{abstract}
\vspace*{.5cm}

\noindent Abstract: Power corrections for virtual Compton
scattering at leading twist are determined at operator level. From
the complete off-cone representation of the twist-2 Compton
operator integral representations for the trace, antisymmetric and
symmetric part of that operator are derived. The operator valued
invariant functions are written in terms of iterated operators and
may lead to interrelations. For matrix elements they go over into
relations for generalized parton distributions. -- Reducing to the
$s$-channel relevant part one gets operator pre-forms of the
Wandzura-Wilczek and the (target mass corrected) Callan-Gross
relations whose structure is exactly the same as known from the
case of deep inelastic scattering; taking non-forward matrix
elements one reproduces earlier results \cite{GRE04} for the
absorptive part of the virtual Compton amplitude. -- All these
relations, obtained without any approximation or using equations
of motion, are determined solely by the twist-2 structure of the
underlying operator and, therefore, are purely of geometric
origin.

\vspace*{0.5cm}


\end{abstract}

\maketitle

\section{Introduction}

Virtual Compton scattering is an excellent example for sharpening our
understanding of hadron physics both theoretically and experimentally.
Already in the early days of QCD, studying deep inelastic scattering (DIS),
many remarkable results were obtained either using simple, approximate
models, also phenomenological ones, or applying general quantum field
theoretical methods. Among these results are the well-known Callan-Gross
(CG) \cite{CG69} and the Wandzura-Wilczek (WW) relations \cite{WW77} for
the structure functions in the unpolarized and polarized DIS, respectively,
and the enlightenment of the influence of target masses \cite{N73,BE76,GP76}.
The extension of these results to deeply virtual Compton scattering (DVCS),
requiring the consideration of non-forward matrix elements of the Compton
operator, appeared to be more difficult -- not only because of the harder
computations but also for conceptual reasons. The latter are connected
with the necessary use of generalized (multi-parameter) distribution
amplitudes -- usually called generalized parton distributions (GPD) --
and with their twist decomposition.

As is well known, the twist decomposition of non-scalar non-local
QCD operators, especially off the light-cone, is highly
nontrivial. Moreover, it suffers from the fact that the notion of
twist is not unique: twist is defined either group theoretically
for the non-local operators and denoted as geometric twist
\cite{GLR99} by extending the original definition for local
operators \cite{GT71} or, differently, it is phenomenological
determined for the individual kinematic invariants of the complete
scattering process \cite{JJ91} counting powers of $1/Q^2$. Both
definitions coincide only at leading order. In the past, investigating 
the group theoretic approach many of its general aspects have been
discovered and generically applied to the relevant light-cone
dominated hard scattering processes. Within that frame the various
higher twist GPDs are related to matrix elements of non-local
operators of higher (geometric) twist.

Recently, studying the target mass corrections of the virtual
non-forward Compton scattering at leading twist a generalization
of the WW- and the (target mass corrected) CG-relations has been
found which have exactly the same formal structure as in the
forward case \cite{GRE04}. This result has been obtained using
only the complete geometric twist-2 structure off-cone, i.e.,
without applying the equations of motion. Analogous observations
have been made earlier when -- extending the consideration of deep
inelastic scattering at leading twist \cite{BR00} -- the parton
distributions at higher twist \cite{GL01}, deep inelastic
scattering with an additional meson in the final state
\cite{BEGR02}, diffractive scattering \cite{BR01} as well as the
behaviour of vector meson distribution amplitudes \cite{L01} had
been considered. These rather interesting results led to the
conjecture \cite{GRE04} that it should be the outcome of the very
structure of the twist-2 Compton operator itself. If so, the
observed relations should follow from matrix elements of
appropriate operator relations and, moreover, would hold also for
other light-cone dominated processes, e.g., with other particles
in the final state which, in fact, is of phenomenological
importance.

The aim of the present paper is to extract the corresponding
operator relations. Thereby, the above mentioned processes are to
be distinguished insofar that in the case of $s$-channel processes, 
like DIS or DVCS, besides the investigation of the entire operator 
expression  -- whose matrix elements correspond to the scattering 
amplitudes -- the absorptive part can be discussed separately. Of course, 
also the amplitudes of light-cone dominated $t$-channel processes, 
like electro-production of hadrons and hadron wave functions, 
are determined by matrix elements of the entire operator expressions. 
However, in that case a separation of the absorptive part at operator 
level remains an open problem, at least presently. In fact, we first
consider the structure of the entire (twist-2) Compton operator and its 
power behavior. Thereafter, we concentrate on that expression which is 
relevant for the $s$-channel absorptive part and we derive its general
operator relations. 

In the next Section, for the sake of notation, we remind the generic
phenomenological situations where the Compton operator appears and introduce
the appropriate approximations as well as point to its geometric twist 
decomposition off-cone.
After that, in three separate Sections we consider the trace part
${\widehat T}^\twz_{\rm trace}(q)$, the antisymmetric part
${\widehat T}^\twz_{[\mu\nu]}(q)$ and the symmetric part
${\widehat T}^\twz_{\{\mu\nu\}}(q)$ of the Compton operator, respectively.
By studying the trace part we already introduce the general procedure of
the investigation. The explicit expressions of the corresponding Fourier
transforms are taking over from our previous work \cite{GRE04}. Then, adapting
the construction of corresponding sets of generalized distribution amplitudes
${\Phi}^{(i)}_a$ of Ref.~\cite{GRE04}, we introduce appropriate sets of
iteratively defined operators ${\cal O}^{(i)}_\alpha$ of order $i$ which
allow to obtain a representation of the various parts of the Compton
operator in terms of independent kinematic invariants. At this stage also
some observations concerning relations for the entire Compton operator are 
made. Then, we consider the operator version of the absorptive parts, denoted by
${\rm Im}\,{\widehat T}^\twz_{\rm trace}(q)$,
${\rm Im}\,{\widehat T}^\twz_{\{\mu\nu\}}(q)$ and
${\rm Im}\,{\widehat T}^\twz_{[\mu\nu]}(q)$, respectively.
We introduce an analogue of the Nachtmann variable together with
a scaling variable which allows to rewrite these expressions
as integrals over appropriate operator valued structure functions.
The latter are completely analogous to the well-known structure functions
occurring in DIS and their generalization to the non-forward case.
In the final, concluding Section we summarize, point to open questions and
comment on related work.

\section{Brief review of prerequisites and relation to previous work}

First, let us present two different generic processes
(at leading twist-2) where the Compton operator appears and remember 
its amplitudes and its kinematics (spin dependence suppressed):
\\
For the $s$-channel we take virtual Compton scattering,
$\gamma^*(q_1) + h(P_1) \rightarrow \gamma^*(q_2) + h(P_2)$, whose
amplitude reads
\begin{eqnarray}
\label{CA_nonf}
^s T^\twz_{\mu\nu}(q; P_i)
 = \langle P_2|
 \Big[\hbox{\large$\int$} \d^4 \! x \; \e^{\im qx}\;
 {\widehat T}^\twz_{\mu\nu}(x)\Big]
 |P_1\rangle \; ,
\end{eqnarray}
where $P_1$ and $P_2$ ($q_1$ and $q_2$) are the momenta of the incoming and outgoing 
hadrons (virtual photons), respectively, $q=(q_2+q_1)/2$,  $p_+ =P_1 + P_2$
and $p_- = P_2 - P_1 = q_1 - q_2\, $ denotes the momentum
transfer.
By crossing, in the $t$-channel, we get electro production of
hadron pairs, $\gamma^*(q_1) + \gamma^*(q_2) \rightarrow h(P_1) + h(P_2)$,
whose amplitude reads
\begin{eqnarray}
\label{pair}
^t T^\twz_{\mu\nu}(q; P_i)
 = \langle \; 0 \;|
 \Big[\hbox{\large$\int$} \d^4 \! x \; \e^{\im qx}\; {\widehat T}^\twz_{\mu\nu}(x)\Big]
 |P_1, P_2\rangle \; ,
\end{eqnarray}
where $q_1$ and $q_2$ are the momenta of the incoming virtual photons
and $P_1$ and $P_2$ are the momenta of the outgoing hadrons; again
$p_\pm = P_2 \pm P_1$ but $q = (q_1 - q_2)/2 $.

Both processes are considered in the generalized Bjorken region,
\vspace*{-.1cm}
\begin{eqnarray}
& q p_+ \longrightarrow \infty,
 \qquad
 Q^2 = -\, q^2 \longrightarrow \infty,&\\
 {\rm with}\quad & x_{\rm Bj}={Q^2}/{qp_+}
\quad{\rm and} \quad \eta ={qp_-}/{qp_+}& \quad {\rm fix}.
\end{eqnarray}
Thereby, the Compton operator in $x$-space is defined by
\begin{eqnarray}
\label{int_input}
 {\widehat T}_{\mu\nu}(x)
 &\equiv& \im\,
 R\,T \kle{\,J_\mu\kln{{x}/{2}}\,J_\nu\kln{-{x}/{2}\,}{\cal S}\,} \,,
\end{eqnarray}
where $J_\mu$ is the electromagnetic current, $R$ denotes the
renormalization procedure and $\cal S$ is the (renormalized)
$S-$matrix. In the generalized Bjorken region, the physical
processes are dominated by the singularities of the Compton
operator on the light-cone and, therefore, the operator product
expansion can be applied. A general study of it, using the
non-local light-cone expansion \cite{AS,MRGHD}, has been given,
e.g., in Refs.~\cite{BGR99,GLR01}, cf.~also~Refs.~\cite{XJ,RAD}.
In leading order the Compton operator simply reads
\begin{eqnarray}
\label{CA_nonf1}
 {\widehat T}_{\mu\nu}(x)
 &\approx&
 \frac{1}{2\,\pi^2\kln{ x^2 - \im \epsilon}^2}\,
 \KLn{{S_{\mu\nu|}}^{\alpha\beta}\,
 x_\alpha\, {\cal O}_\beta\kln{\kappa x, -\kappa x}
 +\,{\epsilon_{\mu\nu}}^{\alpha\beta}
 x_\alpha\, {\cal O}_{5\,\beta}\kln{\kappa x, -\kappa x}} \,,
\end{eqnarray}
where the tensor $S_{\mu\nu|\alpha\beta} \equiv
S_{\mu\alpha\nu\beta} = g_{\mu\alpha} g_{\nu\beta} + g_{\mu\beta}
g_{\nu\alpha} -  g_{\mu\nu} g_{\alpha\beta} $
is symmetric in $\mu\nu$ and $\alpha\beta$ and $\kappa = 1/2$.
(In order to omit the subtleties of operator mixing, the non-singlet
case will be taken also suppressing the flavor structure.)

The hermitean, chiral-even vector and axial vector operators,
 ${\cal O}_{\alpha}(\kappa_1 x,\kappa_2 x)$ and
 ${\cal O}_{5\,\alpha}(\kappa_1 x,\kappa_2 x)$,
respectively, are renormalized on the light-cone, ${\tilde x}^2 =
0$, first and then, in order to allow for an appropriate target
mass expansion, extended off-cone by replacing $\tilde x
\rightarrow x$. They are given by
\begin{eqnarray}
\label{str_oprab}
 {\cal O}_{\alpha}(\kappa_1 x,\kappa_2 x)
 &=&
 \im \big(O_{\alpha}(\kappa_1 x,\kappa_2 x)
 - O_{\alpha}(\kappa_2 x,\kappa_1 x)\big),
 \\
\label{str_oprbb}
 {\cal O}_{5\,\alpha}(\kappa_1 x,\kappa_2 x)
 &=&
 O_{5\,\alpha}(\kappa_1 x,\kappa_2 x)
 +
 O_{5\,\alpha}(\kappa_2 x,\kappa_1 x),
  \end{eqnarray}
\vspace*{-.4cm}
 with
\begin{eqnarray}
 O_{\alpha}(\kappa_1 x,\kappa_2 x)
 &:\,=&
 R \, T \Kle{:\bar\psi\kln{\kappa_1 \tilde x} \gamma_\alpha\,
 U\!\kln{\kappa_1 \tilde x,\kappa_2 \tilde x}\,
 \psi\kln{\kappa_2 \tilde x}\!:{\cal S}}\big|_{\tilde x \rightarrow x},
 \\
 O_{5\,\alpha}(\kappa_1 x,\kappa_2 x)
 &:\,=&
 R \, T \Kle{:\bar\psi\kln{\kappa_1 \tilde x}  \gamma_5 \gamma_\alpha\,
 U\!\kln{\kappa_1 \tilde x,\kappa_2 \tilde x}\,
 \psi\kln{\kappa_2 \tilde x}\!:{\cal S}}\big|_{\tilde x \rightarrow x}.
 \end{eqnarray}
The time-ordered phase factors, will be omitted in the following,
i.e., the Schwinger-Fock gauge, $x^\mu A_\mu(x)=0$, will be
assumed. These definitions are chosen such that the generalized
parton distributions (GPD) which follow by taking matrix elements
would be real (analytic) functions.

Now, in the $s$-channel the absorptive part of the amplitude,
${\rm Im}\;^s T^\twz_{\mu\nu}(q; P_i)$, may be considered which,
after Fourier transformation of expression (\ref{CA_nonf1}),
results from the imaginary part of the quark propagator. This
observation survives at the operator level. Therefore, we are able
to construct an operator, denoted by
 ${\rm Im}\, \widehat T_{\mu\nu}^\twz(q)$, whose matrix elements
 are the absorptive part of the amplitudes in the $s$-channel,
\begin{eqnarray}
\label{imaginar}
 \langle P_2 | {\rm Im}\; \widehat T^\twz_{\mu\nu}(q)
 |P_1\rangle =
 {\rm Im}\;^s T^\twz_{\mu\nu}(q; P_i)\,.
\end{eqnarray}
In fact, for that part of the amplitude of the virtual Compton
scattering the WW- and (mass corrected) CG-relations were derived
in Ref.~\cite{GRE04}. They will be formulated here as operator
relations. However, for $t$-channel processes the complete
amplitude has to be considered. Despite of this, also in these
cases, especially for meson distribution amplitudes as well as
electro production of meson pairs, WW-like relations have been
derived, either using equations of motion \cite{BB95,BBKT,AT01,RW01} 
or using the notion of geometric twist alone \cite{L01}.

Obviously, the operators (\ref{str_oprab}) and (\ref{str_oprbb})
do not have definite twist -- neither on-cone nor off-cone. The
bi-local off-cone (axial) vector operators $O_{(5)\alpha}(\kappa_1
x,\kappa_2 x)$ contain contributions of any geometric twist $\tau
= 2, 3, \ldots \infty$. The corresponding operators of definite
twist recently have been determined in $x$--space \cite{EG03}; by
restriction on the light-cone one gets operators of geometric
twist $\tau = 2,3,4$ which are known already from
Ref.~\cite{GLR99}. The general procedure which allowed to
generalize the notion of geometric twist for the local operators
\cite{GT71} to the case of non-local ones \cite{GLR99}, cf.~also
\cite{GLR01}, consisted in expanding the non-local operators into
a series of local ones, then decomposing these local operators
with respect to irreducible representations of the Lorentz group
and ordering them into towers of related operators of equal twist and,
finally, summing up these towers into separate non-local operators
again. 

In the rest of the paper we restrict ourselves to the off-cone Compton 
operator of geometric twist 2,
\begin{align}
\label{CO2}
  \widehat T_{\mu\nu}^{\,\twz}(x)=
  \frac{1}{2\pi^2(x^2-i\epsilon)^2}&
  \Big({S_{\mu\nu}}^{\alpha\beta}
   x_\alpha {\cal O}_\beta^{\twz} \kln{\kappa x, -\kappa x}
   +{\epsilon_{\mu\nu}}^{\alpha\beta}
   x_\alpha {\cal O}_{5\beta}^{\twz}\kln{\kappa x, -\kappa x}
   \Big),
\end{align}
 \vspace*{-.3cm} {with \cite{GLR01}}
\begin{align}
 \label{NLO}
 O^{\rm tw 2}_{(5)\alpha} \kln{\kappa x,-\kappa x}
 &=
 \sqrt{\pi} \,\pd_\alpha \int_0^1 d\tau
 \int \D^4 u\, O_{(5)\mu}(u) \left\{x^\mu (2 + \kappa\pd_\kappa)
  - \hbox{\large$\frac{1}{2}$}\,\im\kappa \tau\, u^\mu x^2\right\}
 (3 + \kappa\pd_\kappa){\cal H}_1(\tau u;\kappa x)\,,
\\
 {\cal H}_1(\tau u;\kappa x)
 &=\Big(\kappa \tau \sqrt{(ux)^2 - u^2 x^2}\Big)^{-1/2}
 J_{1/2}\Big(\hbox{\large$\frac{\kappa \tau}{2}$} \sqrt{(ux)^2-u^2 x^2}\Big)
 e^{\im \kappa \tau(ux)/2 }\,,
  \nonumber
\end{align}
 and, instead of performing matrix elements, we continue the
investigation of the operator itself. Thereby, as a prerequisite,
use is made of a formal Fourier transformation 
$  O_{(5)\alpha} \kln{\kappa x,-\kappa x} = 
\int \D^4 u\, O_{(5)\mu}(u)\, e^{\im \kappa (xu)};$
for simplicity of notation, the measure is written as 
$\D^4 u = \d^4 u /(2\pi)^4$. The appearance of the non-decomposed 
operator ${\cal O}_\alpha\!\kln{u}$ looks strange but,
in fact, the remaining part of the integrand  by construction acts as a
projector onto its twist-2 part.

Without further approximations, the operator (\ref{CO2}) is used in 
our geometrically based derivation of generalized WW-relations -- 
as well as other ones -- which relate different kinematic structures 
of that operator only. Let us emphasize again that the notion of 
geometric twist differs from the dynamical one \cite{JJ91} which is
conventionally used in the derivation of generalized WW-relations 
in DVCS by relating dynamical twist-2 and twist-3 contributions 
through the equations of motion \cite{BM00,APT00,RW00}. (Unfortunately,
the different notions of `twist' -- geometrical and dynamical are not 
the only ones -- are often a source of confusion and misunderstanding.)

A general, straightforward procedure of determining the Fourier
transform of QCD operators of definite twist, when multiplied by
the (leading part of the) propagator, has been introduced in
Ref.~\cite{EG02} for scalar operators. In the recent work
\cite{GRE04} it has been applied to the twist-2 (axial) vector
operators; these results will be used here. In addition, all
computations to be made here, are similar to those which have been
done in the case of non-forward scattering and, therefore, will
not be repeated in detail; for these explicit computations the
reader is referred to \cite{GRE04}.

In the following we consider the various irreducible tensor
components separately. Thereby, for the clarity of presentation, 
we use the same ordering of the material within each separate Section, 
namely, we consider first (A) the structure of the entire Compton 
operator, then (B) we restrict to the operator expression corresponding
to the $s$-channel absorptive part as well as finally 
(C) we derive the resulting operator relations of the latter.


\section{Trace part of the twist-2 Compton operator}           %

Let us demonstrate the procedure, being sketched in the
previous section, for the simplest example -- the trace part of the
twist-2 Compton operator. Thereby, we also discuss some
phenomenological aspects of the various structures. \\

\noindent {\sf A) Trace part of the entire operator:} \\
The trace
of expression (\ref{CO2}) already has been computed in Ref.~\cite{GRE04}; it
reads
\begin{eqnarray}
\label{FTrace}
\widehat T^{\twz}_{\rm trace}(q)
 &=&
 -\,2\int \frac{\d^4 \! x}{2\pi^2} \;
  \frac{\e^{\im qx}\, x^{\alpha} }{\kln{x^2 - \im\epsilon}^2} \;
 \im\Big(O_{\alpha}^{\twz} \kln{\kappa x, - \kappa x}
 -
 O_{\alpha}^{\twz} \kln{- \kappa x, \kappa x}\Big)
\nonumber\\
&=&
 -\,2\int_0^1 \d\tau \; \int \D^4 u  \;
 {\cal O}_\alpha\!\kln{u} \; \frac{q^2\kln{q^\alpha +
 \kappa\tau u^\alpha}}{\big[\kln{q+\kappa\tau u}^2 +
 \im\epsilon\big]^2}\;,
\end{eqnarray}
where ${\cal O}_\alpha\!\kln{u} = - \,{\cal O}_\alpha\!\kln{-u}$.
Performing the $\tau$--integration partially one obtains
\begin{align}
\widehat T^{\twz}_{\rm trace}(q)
 =
 \int \!\frac{\D^4 u}{\kappa^4}  \,
 {\cal O}_\alpha\!\kln{\frac{u}{\kappa}}
  \frac{- q^2}{[(qu)^2 - q^2u^2]}\,
  \bigg\{&
  \frac{u^\alpha(q^2 + qu) - q^\alpha(qu + u^2)}{R(1) + \im \epsilon}
  - \frac{u^\alpha q^2 - q^\alpha(qu)}{R(0) + \im \epsilon}
  +\!\int_0^1\!\! d\tau
  \frac{u^\alpha(qu) - q^\alpha u^2}{R(\tau) + \im \epsilon}\bigg\},
  \nonumber
\end{align}
where $R(\tau) \equiv (q + \tau u)^2 $.
Introducing the set of antisymmetric
operators ${\cal O}^{(i)}_\alpha\!\kln{u}$ of order $i$
as follows,
\begin{eqnarray}
 {\cal O}^{(0)}_\alpha\!\kln{u}
 \equiv
 {\cal O}_\alpha\!\kln{u}
 \quad {\rm and} \quad
 {\cal O}^{(i)}_\alpha\!\kln{u}
 \equiv
 \int_0^1 \frac{d\tau_1}{\tau_1^4}
 \cdots\int_0^1 \frac{d\tau_i}{\tau_i^4}\,
 {\cal O}_\alpha\!\kln{\frac{u}{\tau_1\ldots\tau_i}}
 \quad {\rm for} \quad i\geq 1,
 \label{GDA}
\end{eqnarray}
then, after scaling the remaining variables $\tau u$, we obtain
\begin{align}
\widehat T^{\twz}_{\rm trace}(q)
 =
 - \int \frac{\D^4 u}{\kappa^4}   \;
 \frac{q^2}{[(qu)^2 - q^2u^2]}\,\bigg\{
 & \left(u^\alpha q^2 - q^\alpha (qu) \right)
 {\cal O}^{(0)}_\alpha\!\kln{\frac{u}{\kappa}}
 \bigg[\frac{1}{R(1) + i\epsilon}- \frac{1}{R(0) + i\epsilon}
 \bigg]
 \nonumber\\
 + & \left( u^\alpha (qu) - q^\alpha u^2 \right)
   \bigg[\;{\cal O}^{(0)}_\alpha\!\kln{\frac{u}{\kappa}}
   + {\cal O}^{(1)}_\alpha\!\kln{\frac{u}{\kappa}}\bigg]
   \frac{1}{R(1)+i\epsilon}
 \bigg\}.
  \label{Ftrace0}
 \end{align}

This expression has an interesting structure. First, let us remark
 that in (\ref{Ftrace0}) the iterated operators ${\cal O}^{(i)}_\alpha$
 occur instead of iterated distribution amplitudes which, of course,
 reappear if appropriate matrix elements are built. (The same
 occurs later on when the (anti)symmetric part of the entire Compton
 operator is considered.) Obviously, the appearance of the iterated
 operators (\ref{GDA}) is a
consequence of the whole structure of the twist-2 Compton
operator. Therefore, it contains essential elements which are
typical for the derivation of WW-relations. In addition, the
operator combinations of both kinematical structures of
(\ref{Ftrace0}) are very simply related (if the $1/R(0)$-term is
ignored). Let us consider that expression more detailed since it
is of technical relevance in our further consideration.

First, introducing $u^\top_\alpha$ as that component of $u_\alpha$
which is transversal with respect to the (external) momentum $q_\alpha$,
\begin{eqnarray}
     u^\top_\alpha &=& g^\top_{\alpha\beta} u^\beta,
     \qquad
     g^\top_{\alpha\beta} = g_{\alpha\beta} - q_\alpha q_\beta / q^2 \\
     (u^\top)^2 &=& [(qu)^2 - q^2u^2]\; / \left(- q^2\right),
\end{eqnarray}
we observe that the two independent kinematic structures can be rewritten as follows:
   \begin{eqnarray}
    v_\alpha  &\equiv&  u_\alpha\, q^2 - q_\alpha (qu)  = q^2\, u^\top_\alpha , \\
    w_\alpha  &\equiv&  u_\alpha (qu) - q_\alpha\, u^2
      =  u^\top_\alpha (qu) - q_\alpha (u^\top)^2.
    \end{eqnarray}
Due to the `dual' structure of $w$ and $v$ similar relations with
$w_\alpha = - u^2 q^\bot_\alpha$ result if
we introduce $q^\bot_\alpha = q_\alpha - u_\alpha (qu)/u^2$.
In addition, the following relations hold
\begin{eqnarray}
     vq = 0 = wu, \qquad
     wq = - vu = - q^2(u^\top)^2,
     \qquad vw  = (u^\top)^2 \left[ -q^2/(qu)\right].
\end{eqnarray}

At this stage we should mention that the iterated operators
 ${\cal O}^{(i)}_\alpha\!\kln{{u}}$ implicitly contain the full 
 kinematical structure of any possible hard QCD process.
 This becomes obvious when the matrix element of the operator
 $\,O^{\rm tw 2}_{\alpha} \kln{ x,- x}$ is decomposed
 into its kinematical independent amplitudes $f_a (x, P_i )$,
\begin{eqnarray}
  \langle P_2,S_2 |\,O^{\rm tw 2}_{\alpha} \kln{ x, -x}|P_1,S_1 \rangle
  =
  \hbox{$\sum_a {\cal  K}_\alpha^a$}\,(P_i, S_i)\,  f_a (x, P_i )\,,
  \label{gda0}
 \end{eqnarray}
where ${\cal  K}_\alpha^a(P_i, S_i)$ are the independent form
factors. An analogous decomposition appears for ${\cal
O}^{(i)}_\alpha\!\kln{{u}}$.

Furthermore, let us decompose $1/R$ into its partial fractions by
rewriting
\begin{eqnarray}
 \label{IM}
 \frac{1}{R(\tau)+ \im\epsilon}=
 \frac{1}{2\sqrt{(qu)^2 - q^2u^2}}
 \left(
   \frac{1}{\tau - \tilde\xi_+ + \im \epsilon}
 - \frac{1}{\tau - \tilde\xi_- - \im \epsilon}
 \right),
\end{eqnarray}
with
\begin{eqnarray}
\label{xi}
    \tilde\xi_\pm
    = \frac{-\, q^2}{qu \pm \sqrt{(qu)^2 - q^2 u^2}}
    = \frac{-\, qu \pm \sqrt{(qu)^2 - q^2u^2}}{u^2}
    \,.
\end{eqnarray}
Obviously, these two equivalent forms of $\tilde\xi_\pm$ show an
analogous structural similarity w.r.t. $u$ and $q$ as has been
observed for $v$ and $w$. Let us also remark that due to the
antisymmetry of the operators ${\cal O}^{(i)}_\alpha\!\kln{u}$
only the symmetric and antisymmetric combinations of the
$1/R$-terms have to be taken when multiplying the kinematic
structures $v$ and $w$, respectively.

Now, let us restrict formally the support of ${\cal O}(u)$ to
regions with $u^2 > 0$ which, in accordance with
Ref.~\cite{GRE04}, makes sense at least if we have in mind
$s$-channel processes. Then, we can scale the variable $u$ (having
mass dimension 1, ${\rm dim}\, u \equiv [u]$) by the dimensionless
`radius' $t$, $0 \leq t \leq \infty$, times a unit vector $\hat
p\,$ of mass dimension 1,
\begin{eqnarray}
u = t\,{\hat p}\,, \qquad t = \sqrt{u^2/{\hat p}^2}\,, \qquad
{\hat p} = \hat n\,[u]\,, \qquad \hat n ={u}/{\sqrt{u^2}}\,,
\qquad {\hat n}^2 = 1\,.
\end{eqnarray}
Then, also $\tilde\xi_\pm$ is simply scaled leading to a (dimensionless)
analogue $\xi_\pm$ of the Nachtmann variable \cite{N73} in terms
of a Bjorken-like variable $x$:
\begin{eqnarray}
\tilde\xi_\pm = \xi_\pm/ t \qquad {\rm with}\qquad \xi_\pm
 = \frac{x}{1 \pm \sqrt{1 + x^2 {\hat p}^2 / Q^2}}
\qquad {\rm and}\qquad
 x \equiv \frac{Q^2}{q\hat p} \,.
\end{eqnarray}
Furthermore, the integration over $u$ gets an integration over the
radius $0\leq t \leq \infty$ and over the 4-dimensional unit
two-shell hyperboloid ${\cal H}^3$.
\\

\noindent
{\sf B) Restriction to the imaginary part}\\
Next, we restrict to the operator expression
 ${\rm Im}\, \widehat T_{\mu\nu}^\twz(q)$
 which describes possible absorptive parts in operator form, i.e.,
 that part which results from the imaginary part of the quark
 propagator.
 Thereby, the use of the $t$-variable, despite probably not helpful
when considering the entire operator (\ref{Ftrace0}), is essential
for the consideration of its imaginary part. Due to the overall
factor $q^2$ in expresseion (\ref{Ftrace0}) that imaginary part results 
from $1/[R(1)+ i\epsilon]$ only. Because
of Eq.~(\ref{IM}), taken at $\tau = 1$, we simply get
\begin{eqnarray}
 \label{imR}
 {\mathrm{Im}} \,\frac{1}{R(1)+ \im\epsilon}=
 -\frac{\pi}{2}\frac{1}{\sqrt{(qu)^2 - q^2u^2}}\,
 \big[ \delta(1 - \tilde\xi_+ ) + \delta(1 - \tilde\xi_- ) \big]\,.
\end{eqnarray}
Obviously, it holds
\begin{eqnarray}
     \left(qu + u^2 \right) \delta(1 - \tilde\xi_\pm)
    = \pm\,\left[\sqrt{(qu)^2 - q^2u^2}\right] \delta(1 - \tilde\xi_\pm)
    = -\left( q^2 + qu\right) \delta(1 - \tilde\xi_\pm)
    \label{+}\,;
    \end{eqnarray}
these equalities are essential and will be used to simplify some of the 
resulting expressions below.

In fact, the imaginary part of expression (\ref{Ftrace0}) gets
\begin{align}
 {\mathrm{Im}}\, \widehat T^\pm_{\rm trace}\kln{q}
 =&~
 \frac{\pi}{2}\int \d t^2\int \frac{\D^4 u}{\kappa^4}
 \; \delta(u^2/[u]^2 - t^2 )\; \delta\big(1 - \tilde\xi_\pm\big)
 \frac{- q^2}{[(qu)^2 - q^2 u^2]^{3/2}}\,
 \times
\nonumber\\
  & \qquad \left\{ \left[
  (qu + u^2)\,q^\alpha{\cal O}^{(0)}_\alpha\!\kln{\frac{u}{\kappa}}
       + u^2\,q^\alpha{\cal O}^{(1)}_\alpha\!\kln{\frac{u}{\kappa}}\right]
  -
  \left[(q^2 + qu)\,u^\alpha{\cal O}^{(0)}_\alpha\!\kln{\frac{u}{\kappa}}
            + (qu)\,u^\alpha{\cal O}^{(1)}_\alpha\!\kln{\frac{u}{\kappa}}
  \right]
  \right\}
\nonumber\\
  =&~
  \pi
  \int \frac{\D^4 {\hat p}}{\kappa^4}\,\delta({\hat p}^2 - 1)
  \int \d t\,t^4\;\delta\big(t - \xi_\pm\big)
  \frac{- q^2}{t^3[(q{\hat p})^2 - q^2 {\hat p}^2]^{3/2}}\,
  \times
\nonumber\\
  &
  \qquad \bigg\{\left[
  \pm t\,[(q{\hat p})^2 - q^2 {\hat p}^2]^{1/2}\,
  q^\alpha{\cal O}^{(0)}_\alpha\Big(\frac{t{\hat p}}{\kappa}\Big)
  + t^2\,{\hat p}^2\,
  q^\alpha{\cal O}^{(1)}_\alpha\Big(\frac{t{\hat p}}{\kappa}\Big)\right]
\nonumber\\
  & \qquad -
  \left[
  \mp t^2\,[(q{\hat p})^2 - q^2 {\hat p}^2]^{1/2}\,
  {\hat p}^\alpha{\cal O}^{(0)}_\alpha\Big(\frac{t{\hat p}}{\kappa}\Big)
  + t^2\,(q{\hat p})\,
  {\hat p}^\alpha{\cal O}^{(1)}_\alpha\Big(\frac{t{\hat p}}{\kappa}\Big)
  \right]
  \bigg\}
\nonumber\\
  =&~ \pi
  \int_{{\cal H}^3} \frac{\D^4 \hat p}{\kappa^4}\,\delta({\hat p}^2 - 1)
  \int_0^\infty \d t \;\delta\big(t - \xi_\pm\big) %
  \frac{x}{1+x^2{\hat p}^2/Q^2}\,
  \Bigg\{
  \pm\, \left(1 + t\,x\,\frac{{\hat p}^2}{Q^2}\right)\;
  \frac{q^\alpha}{q{\hat p}}\;
  {\Omega}_\alpha^{(0)}\Big(t,\frac{\hat p}{\kappa}\Big)
\nonumber\\
&
 \qquad\quad +
 x\,\frac{{\hat p}^2}{Q^2}
 \left( \frac{q^\alpha}{q{\hat p}}
 -
 \frac{{\hat p}^\alpha}{{\hat p}^2}
 \right)
 \left[\mp\,t\;
 {\Omega}_\alpha^{(0)}\Big(t,\frac{\hat p}{\kappa}\Big)
 + \frac{1}{\sqrt{1+x^2{\hat p}^2/Q^2}} \;
 {\Omega}_\alpha^{(1)}\Big(t,\frac{\hat p}{\kappa}\Big)
 \right] \Bigg\}\,;
\label{Trop}
\end{align}
here and in the following, we maintain ${\hat p}^2 = 1$ in order to remaind
its mass dimension without denoting it explicitly. In deriving expression
(\ref{Trop}), Eq.~(\ref{+}) has been used repeatedly and
 $t^2\,{\cal O}^{(0)}_\alpha\!\kln{u},\; t^3\,{\cal O}^{(1)}_\alpha\!\kln{u}$
and so on has been replaced by
\begin{eqnarray}
\label{Phi_tr}
 {\Omega}_\alpha^{(0)}(t,\hat p)
 \equiv
 t^2\,{\cal O}^{(0)}_\alpha\!\kln{u} ,
 \quad
 {\Omega}_\alpha^{(1)}(t,\hat p)
 \equiv
 t^3\,{\cal O}^{(1)}_\alpha\!\kln{u}
 = t^3 \int_0^1 \frac{\d\tau}{\tau^4}\;
   {\cal O}_\alpha\Big(\frac{t\hat p}{\tau}\Big)
 = \int_t^\infty\!\! dy \,y^2\,{\cal O}_\alpha\!\left(y \hat p\right)
 = \int_t^\infty\!\! dy \,{\Omega}_\alpha^{(0)}(y,\hat p),
\end{eqnarray}
and so on.

Now, taking into account the following relations
\begin{eqnarray}
    \xi_\pm(x,Q^2) &= &
    \frac{x}{1 \pm \sqrt{1+x^2 {\hat p}^2/Q^2}}
    =
    - \frac{Q^2}{{\hat p}^2}\,
    \frac{1 \mp \sqrt{1+x^2 {\hat p}^2/Q^2}}{x},
    \label{rel1}\\
    x\frac{\partial}{\partial x}\left(
    \frac{x}{[1+x^2 {\hat p}^2/Q^2]^{1/2}}\right)
    &=&
    \frac{x}{[1+x^2 {\hat p}^2/Q^2]^{3/2}}\,,
    \label{rel2}\\
    x\frac{\partial}{\partial x} \xi_\pm
    &=&
    \pm  \frac{\xi_\pm}{[1+x^2 {\hat p}^2/Q^2]^{1/2}}\,,
    \label{rel3}\\
    x\frac{\partial}{\partial x}{\Omega}_{\alpha}^{(1)}(\xi_\pm,\hat p)
    &=&
    \mp  \frac{\xi_\pm}{[1+x^2 {\hat p}^2/Q^2]^{1/2}}
    {\Omega}_{\alpha}^{(0)}(\xi_\pm,\hat p)\,,
    \label{rel4}
\end{eqnarray}
and introducing the following operator functions,
\begin{align}
    & {\cal V}_{0\alpha}(\xi_\pm,x;\hat p)
   \equiv
   \frac{x\;
 {\Omega}_{\alpha}^{(0)}(\xi_\pm,\hat p)}{\sqrt{1+x^2{\hat p}^2/Q^2}}    \,,
\label{H0}
    \\
   & {\cal V}_{1\alpha}(\xi_\pm,x;\hat p)
   \equiv
   x\frac{\partial}{\partial x}\left(
   \frac{x\,
   {\Omega}_{\alpha}^{(1)}(\xi_\pm,\hat p)}{\sqrt{1+x^2{\hat p}^2/Q^2}}
   \right)
    =
    \frac{x}{1+x^2 {\hat p}^2/Q^2}\,
    \left[
    \mp\,\xi_\pm\, {\Omega}_{\alpha}^{(0)}(\xi_\pm,\hat p)
    +
    \frac{1}{\sqrt{1+x^2 {\hat p}^2/Q^2}}\;
    {\Omega}_{\alpha}^{(1)}(\xi_\pm,\hat p)
    \right],
\label{H1}
\end{align}
one finally finds for the imaginary part of the trace of the
twist-2 Compton operator:
\begin{eqnarray}
 {\mathrm{Im}}\, {\widehat T}^\twz_{\rm trace}\kln{q}
 &=&
 2 {\pi}
 \int_{{\cal H}^3} \frac{\D^4 \hat p}{\kappa^4}\; \delta({\hat p}^2 - 1) %
 \left\{
 \frac{q_\alpha }{q{\hat p}}\;
 {\cal V}_{0}^{\alpha}\Big(x,\frac{\hat p}{\kappa}\Big)
 +
 x\,\frac{{\hat p}^2}{Q^2}
 \left(
 \frac{q_\alpha}{q{\hat p}} - \frac{{\hat p}_\alpha}{{\hat p}^2}
 \right)
 {\cal V}_{1}^{\alpha}\Big(x,\frac{\hat p}{\kappa}\Big)
 \right\}
 \label{FTraceX}
 \\
 &=&
 2 {\pi}
 \int_{{\cal H}^3} \frac{\D^4 \hat p}{\kappa^4}\; \delta({\hat p}^2 - 1) %
 \left\{
 \frac{q_\alpha }{q{\hat p}}\;
 \bigg[
 {\cal V}_{0}^{\alpha}\Big(x,\frac{\hat p}{\kappa}\Big)
 + \frac{\left({\hat p}^\top\right)^2}{q{\hat p}}
 {\cal V}_{1}^{\alpha}\Big(x,\frac{\hat p}{\kappa}\Big)
 \bigg]
 -  \frac{{\hat p}^\top_\alpha}{q{\hat p}}\;
 {\cal V}_{1}^{\alpha}\Big(x,\frac{\hat p}{\kappa}\Big)
  \right\}
  \label{FTraceX1}
\end{eqnarray}
with ${\cal V}_{i\alpha }(x,\hat p) =
 {\cal V}_{i\alpha }(\xi_+,x;\hat p) +
 {\cal V}_{i\alpha }(\xi_-,x;\hat p),\, i = 0,1$.

\section{Antisymmetric part of the twist-2 Compton operator}
\noindent{\sf A) Entire operator expression} \\
Now, we consider the antisymmetric part of the Compton operator
which is given by \cite{GRE04}
\begin{eqnarray}
 {\widehat T}^{\twz}_{[\mu\nu]}(q)
 &=&
 \epsilon_{\mu\nu}^{~~\alpha\beta}
 \int \frac{\d^4 \! x}{2\pi^2} \;
  \frac{\e^{\im qx}\, x_{\alpha} }{\kln{x^2 - \im\epsilon}^2} \;
 \Big(O_{5\beta}^{\twz} \kln{\kappa x, - \kappa x}
 +
 O_{5\beta}^{\twz} \kln{- \kappa x, \kappa x}\Big)
\nonumber\\
&=&
 \epsilon_{\mu\nu}^{~~\alpha\beta}\,
 \int_0^1 \d\tau \int _0^1 \d\tau' \int \frac{\D^4 u}{\kappa^4}
 \;{\cal O}_{5\rho} \kln{{\frac{u}{\kappa}}} \,
 \partial^\rho_u \kln{ \frac{{q^2}\,q_{\kleo{\alpha}}u_{\okle{\beta}}}
 {[(q + \tau\tau' u)^2 + i\epsilon]^2 }}\,.
\label{OAS1}
\end{eqnarray}
Again, the $\tau$--integrations may be performed partially so that we
end up with $1/[R(1)+ \im \epsilon]$ as well as single and double
integrals over $1/[R(\tau)+ \im \epsilon]$ and
$1/[R(\tau\tau')+ \im \epsilon]$, respectively.
Then, performing that tedious but straightforward calculation and scaling
away the respective $\tau-$dependence, the result finally reads
(cf.~computations in Ref.~\cite{GRE04})
\begin{align}
 {\widehat T}^\twz_{[\mu\nu]}\kln{q}
    =&
    \int \frac{\D^4 u}{\kappa^4} \;
    \frac{-\,q^2\,{\epsilon_{\mu\nu}}^{\alpha\beta}\,q_\alpha}
    {2\,[(qu)^2 - q^2 u^2]}\,\frac{1}{R(1)+ \im \epsilon}\,
    \bigg\{
    {g_{\beta}^\top}^{\rho}\Big[
    (qu+u^2)\,{\cal O}_{5\rho}^{(1)}\Big(\frac{u}{\kappa}\Big)
    +   u^2 \,{\cal O}_{5\rho}^{(2)}\Big(\frac{u}{\kappa}\Big)
    \Big]
    \nonumber\\
    &\qquad
    + u_\beta^\top\,q^\rho \left[
      {\cal O}_{5\rho}^{(0)}\Big(\frac{u}{\kappa}\Big)
    +2\,{\cal O}_{5\rho}^{(1)}\Big(\frac{u}{\kappa}\Big)
    - 3\frac{ qu }{[(qu)^2 - q^2 u^2]}
    \left[
    (qu+u^2)\,{\cal O}_{5\rho}^{(1)}\Big(\frac{u}{\kappa}\Big)
    +   u^2 \,{\cal O}_{5\rho}^{(2)}\Big(\frac{u}{\kappa}\Big)
    \right]
    \right]
    \nonumber\\
    &\qquad
    + u_\beta^\top\,u^\rho \left[
    {\cal O}_{5\rho}^{(0)}\Big(\frac{u}{\kappa}\Big)
    - {\cal O}_{5\rho}^{(2)}\Big(\frac{u}{\kappa}\Big)
    + 3 \frac{qu }{[(qu)^2 - q^2 u^2]}
    \left[
    (q^2 + qu)\,
    {\cal O}_{5\rho}^{(1)}\Big(\frac{u}{\kappa}\Big)
    + (qu)\,
    {\cal O}_{5\rho}^{(2)}\Big(\frac{u}{\kappa}\Big)
    \right]
    \right]
    \bigg\}\,,
    \label{Tas500}
\end{align}
where the set of operators ${\cal O}_{5\mu}^{(i)}(u)$
are defined in the same manner as ${\cal O}_{\mu}^{(i)}(u)$ in
Eqs.~(\ref{GDA}); possible terms proportional to
${1}/[R(0)+ \im \epsilon]$ vanish due to the symmetry of the operators,
${\cal O}^{(i)}_{5\alpha}(u) = {\cal O}^{(i)}_{5\alpha}(-u)$. The
transverse structures are due to the $\epsilon$-tensor.
With the help of $v_\rho= q^2 u^\top_\rho$ and
$w_\rho = (qu)u^\top_\rho - (u^\top_\rho)^2 q_\rho$ we get
another structure,
\begin{align}
{\widehat T}^\twz_{[\mu\nu]}\kln{q}    =&
    \int \frac{\D^4 u}{\kappa^4} \;
    \frac{-\,q^2\,{\epsilon_{\mu\nu}}^{\alpha\beta}\,q_\alpha}
    {2\,[(qu)^2 - q^2 u^2]}\,\frac{1}{R(1)+ \im \epsilon}\,
    \bigg\{
    {g_{\beta}^\top}^{\rho}\Big[
    (qu+u^2)\,{\cal O}_{5\rho}^{(1)}\Big(\frac{u}{\kappa}\Big)
    +   u^2 \,{\cal O}_{5\rho}^{(2)}\Big(\frac{u}{\kappa}\Big)
    \Big]
    \nonumber\\
    &\qquad
    + \frac{u_\beta^\top\,w^\rho}{(qu)^2 - q^2 u^2}
    \left[
    \left((q^2 + qu)\,
    {\cal O}_{5\rho}^{(0)}\Big(\frac{u}{\kappa}\Big)
    + (qu)\,
    {\cal O}_{5\rho}^{(1)}\Big(\frac{u}{\kappa}\Big)
    \right)
    + 2 \left(
    (q^2 + qu)\,
    {\cal O}_{5\rho}^{(1)}\Big(\frac{u}{\kappa}\Big)
    + (qu)\,
    {\cal O}_{5\rho}^{(2)}\Big(\frac{u}{\kappa}\Big)
    \right)
    \right]
    \nonumber\\
    &\qquad
    - \frac{u_\beta^\top\,v^\rho}{(qu)^2 - q^2 u^2}
    \left[
    \left( 
    (qu + u^2)\,{\cal O}_{5\rho}^{(0)}\Big(\frac{u}{\kappa}\Big)
    +     u^2 \,{\cal O}_{5\rho}^{(1)}\Big(\frac{u}{\kappa}\Big)
    \right)
    -
    \left(
    (qu+u^2)\,{\cal O}_{5\rho}^{(1)}\Big(\frac{u}{\kappa}\Big)
    +   u^2 \,{\cal O}_{5\rho}^{(2)}\Big(\frac{u}{\kappa}\Big)
    \right)
    \right]\!
    \bigg\},
    \label{Tas501}
\end{align}
where the kinematics is now classified according to
 ${g_{\beta}^\top}^{\rho}{\cal O}_{5\rho}^{(i)},\;
 v^\rho {\cal O}_{5\rho}^{(i)}$ and $w^\rho {\cal  O}_{5\rho}^{(i)}$.
 Especially in the latter case one observes remarkable structural similarities 
 of these various expressions in terms of the iterated operators. However, 
 relations being similar to those which will be derived now for the imaginary
 part have not been found. 
\bigskip

\noindent {\sf B) Restriction to the imaginary part
${\mathrm{Im}}\, {\widehat T}^\twz_{[\mu\nu]}\kln{q}$ }\\
 Now, let us use in Eq. (\ref{Tas500}) the representation (\ref{imR})
 of the imaginary
part of ${1}/[R(1)+ \im \epsilon]$ and  performing the same steps
as in Eq.~(\ref{Trop}). Namely, introducing the additional
$t-$integration, rewriting $u = t {\hat p}$ and replacing the set
of operators $t^3{\cal O}_{\alpha}^{(i)}(t\, {\hat p})$ by
\begin{eqnarray}
{\Omega}_{5\alpha}^{(0)}(t, {\hat p})
 = t^3 \,{\cal O}_{\alpha}(t\,{\hat p})\,,
 \qquad
 {\Omega}_{5\alpha}^{(i)}(t, {\hat p})
 = \int_t^\infty \frac{\d y}{y} \;
 {\Omega}_{5\alpha}^{(i-1)}(y,\,{\hat p})
 \qquad {\rm for} \qquad
 i\geq 1\,,
\end{eqnarray}
we obtain
\begin{align}
{\mathrm{Im}}\, {\widehat T}^\twz_{[\mu\nu]}\kln{q} =&\;
 {\mathrm{Im}}\, {\widehat T}^+_{[\mu\nu]}\kln{q}
 +
 {\mathrm{Im}}\, {\widehat T}^-_{[\mu\nu]}\kln{q}\,,
 \nonumber\\
 {\mathrm{Im}}\, {\widehat T}^\pm_{[\mu\nu]}\kln{q}
=&
  \frac{\pi}{2}
  \int_{{\cal H}^3} \frac{\D^4 \hat p}{\kappa^4}\,\delta({\hat p}^2 - 1)\,
    \epsilon_{\mu\nu}^{\phantom{\mu\nu}\alpha\beta}\,
    \left(-\frac{x}{\xi_\pm}\,\frac{1}{[1+x^2{\hat p}^2/Q^2]^{3/2}}\right)
    \times
\label{Tas6}
\\
    & \quad
    \Bigg\{
    \frac{q_\alpha g_\beta^\rho}{q\hat p}
    \left[\Big(1 + x\,\xi_\pm\,\frac{{\hat p}^2}{Q^2}\Big)\;
    {{\Omega}}_{5\rho}^{(1)}\Big(\xi_\pm,\frac{\hat p}{\kappa}\Big)
    + x\,\xi_\pm\,\frac{{\hat p}^2}{Q^2}\;
    {{\Omega}}_{5\rho}^{(2)}\Big(\xi_\pm,\frac{\hat p}{\kappa}\Big)
    \right]
\nonumber\\
    & \quad
    + \frac{q_{\alpha}\,{\hat p}_\beta}{q{\hat p}}
    \frac{q^\rho}{q\hat p}
    \left[
    {{\Omega}}_{5\rho}^{(0)}\Big(\xi_\pm,\frac{\hat p}{\kappa}\Big)
    \mp
    \frac{1 - 2\,x\,\xi_\pm\,{\hat p}^2/Q^2}{[1+x^2{\hat p}^2/Q^2]^{1/2}}\;
    {{\Omega}}_{5\rho}^{(1)}\Big(\xi_\pm,\frac{\hat p}{\kappa}\Big)
    -
    \frac{3\,x\,\xi_\pm\,{\hat p}^2/Q^2 }{[1+x^2{\hat p}^2/Q^2]}\;
    {{\Omega}}_{5\rho}^{(2)}\Big(\xi_\pm,\frac{\hat p}{\kappa}\Big)
    \right]
\nonumber\\
    & \quad
    + x\,\xi_\pm\,\frac{{\hat p}^2}{Q^2}
    \frac{q_{\alpha}\,{\hat p}_\beta}{q{\hat p}}
    \frac{{\hat p}^\rho}{{\hat p}^2}
    \left[
    {{\Omega}}_{5\rho}^{(0)}\Big(\xi_\pm,\frac{\hat p}{\kappa}\Big)
    \mp
    \frac{3}{[1+x^2 {\hat p}^2/Q^2]^{1/2}}\;
    {{\Omega}}_{5\rho}^{(1)}\Big(\xi_\pm,\frac{\hat p}{\kappa}\Big)
    +
    \frac{2- x^2 {\hat p}^2/Q^2}{[1+x^2 {\hat p}^2/Q^2]}\;
    {{\Omega}}_{5\rho}^{(2)}\Big(\xi_\pm,\frac{\hat p}{\kappa}\Big)
    \right]\!\!
    \Bigg\}.
    \nonumber
\end{align}
Now, observing Eqs.~(\ref{rel3}) and (\ref{rel4}) together with
\begin{eqnarray}
    x\frac{\partial}{\partial x}\left(
    \frac{x}{\xi_\pm}\frac{1}{[1+x^2 {\hat p}^2/Q^2]^{1/2}}\right)
    &=&
    \frac{x}{\xi_\pm}
    \frac{1\mp [1+ x^2 {\hat p}^2/Q^2]^{1/2}}{[1+x^2 {\hat p}^2/Q^2]^{3/2}}
    =
    - \frac{x^2 {\hat p}^2/Q^2 }{[1+x^2 {\hat p}^2/Q^2]^{3/2}}\,,
    \nonumber
\end{eqnarray}
we get
\begin{align}
{\mathrm{Im}}\, {\widehat T}^\twz_{[\mu\nu]}\kln{q}
    =&
    -\frac{\pi}{2}\int_{{\cal H}^3} \frac{\D^4 \hat p}{\kappa^4}
    \,\delta({\hat p}^2 - 1)\,
    \epsilon_{\mu\nu}^{\phantom{\mu\nu}\alpha\beta}\,
    \Bigg\{
    \frac{q_{\alpha}\,{g_\beta}^\rho}{q{\hat p}}
    \left[
      {{\cal G}}_{1\rho}\Big(x,\frac{\hat p}{\kappa}\Big)
    + {{\cal G}}_{2\rho}\Big(x,\frac{\hat p}{\kappa}\Big)
    \right]
    \nonumber\\
     &\qquad\qquad\qquad\qquad\qquad\qquad
    - \frac{q_{\alpha}\,{\hat p}_\beta}{q{\hat p}}
    \left[
    \frac{q^\rho}{q\hat p}\,
    {{\cal G}}_{2\rho}\Big(x,\frac{\hat p}{\kappa}\Big)
    -
    \frac{{\hat p}^\rho}{Q^2}\,
    {{\cal G}}_{0\rho}\Big(x,\frac{\hat p}{\kappa}\Big)
    \right]
    \Bigg\}\,,
    \label{Tas7}
\end{align}
with ${\cal G}_{i\rho}(x,{\hat p}/{\kappa})
    =
    {\cal G}^+_{i\rho}(x,{\hat p}/{\kappa})
    +
    {\cal G}^-_{i\rho}(x,{\hat p}/{\kappa})
    $
     for
    $
    i = 0,1,2\,,
$ and
\begin{align}
   & {\cal G}^\pm_{1\rho}(x,{\hat p}/{\kappa})
   \equiv
    x\frac{\partial}{\partial x}\left[
    x\frac{\partial}{\partial x}\left(
    \frac{x}{\xi_\pm}
    \frac{{\Omega}_{5\rho}^{(2)}(\xi_\pm,{\hat p}/{\kappa})}
    {[1+x^2 {\hat p}^2/Q^2]^{1/2}}
    \right)\right],
    \label{G_rel1}
    \\
    &  {\cal G}^\pm_{2\rho}(x,{\hat p}/{\kappa})
    \equiv
    - x\frac{\partial^2}{\partial x^2}x\left(
    \frac{x}{\xi_\pm}
    \frac{{\Omega}_{5\rho}^{(2)}(\xi_\pm,{\hat p}/{\kappa})}
    {[1+x^2 {\hat p}^2/Q^2]^{1/2}}
    \right),
    \label{G_rel02}
    \\
    & {\cal G}^\pm_{0\rho}(x,{\hat p}/{\kappa})
    \equiv
    x^2 \frac{\partial^2}{\partial x^2}
    \left( x^2
    \frac{{\Omega}_{5\rho}^{(2)}(\xi_\pm,{\hat p}/{\kappa})}
    {[1+x^2 {\hat p}^2/Q^2]^{1/2}}
    \right).
    \label{G_rel0}
\end{align}
\bigskip

\noindent {\sf C) General operator relations between
${\cal G}^\pm_{i\,\rho}(x,{\hat p}/{\kappa})$} 
\\
 Obviously, Eqs.~(\ref{G_rel1}) -- (\ref{G_rel0}) contain only two independent quantities
which contribute, namely, introducing ${\mathcal F}_\rho(x,{\hat
p}/{\kappa})= {\mathcal F}^+_\rho(x,{\hat p}/{\kappa}) + {\mathcal
F}^-_\rho(x,{\hat p}/{\kappa})$ and ${\mathcal F}^0_\rho(x,{\hat
p}/{\kappa})= \xi_+{\mathcal F}^+_\rho(x,{\hat p}/{\kappa}) +
\xi_-{\mathcal F}^-_\rho(x,{\hat p}/{\kappa})$ with
\begin{eqnarray}
{\mathcal F}^\pm_\rho\Big(x,\frac{\hat p}{\kappa}\Big)
    =
    \frac{x}{\xi_\pm}
    \frac{{\Omega}_{5\rho}^{(2)}(\xi_\pm,{\hat p}/{\kappa})}
    {[1+x^2 {\hat p}^2/Q^2]^{1/2}}\,,
\end{eqnarray}
we may rewrite the relations (\ref{G_rel1}) -- (\ref{G_rel0}) as follows:
\begin{eqnarray}
    {\cal G}_{1\rho}\Big(x,\frac{\hat p}{\kappa}\Big) &=&
    x\frac{\partial}{\partial x}\;
    x\frac{\partial}{\partial x}\;
    {{\cal F}}_\rho\Big(x,\frac{\hat p}{\kappa}\Big)\,,
    \nonumber\\
    {\cal G}_{2\rho}\Big(x,\frac{\hat p}{\kappa}\Big) &=&
    - x\frac{\partial}{\partial x}\;
    \Big(x\frac{\partial}{\partial x} + 1\Big)\,
    {{\cal F}}_\rho\Big(x,\frac{\hat p}{\kappa}\Big)\,,
    \nonumber\\
    {\cal G}_{0\rho}\Big(x,\frac{\hat p}{\kappa}\Big) &=&
    x\frac{\partial}{\partial x}\;
    \Big(x\frac{\partial}{\partial x} - 1\Big)\,
    x\,{{\cal F}}^0_\rho\Big(x,\frac{\hat p}{\kappa}\Big)\,.
\label{G_rel2}
\end{eqnarray}
The first two of these relations are formally identical in their functional
dependence with the corresponding equations in DIS as
given in Ref.~\cite{BT99}. Therefore, they will have the same
general consequences:
Observing
\begin{eqnarray}
    x\frac{\partial}{\partial x}
    {{\cal F}}_\rho\Big(x,\frac{\hat p}{\kappa}\Big)
    = 
    - \int_x^1 \frac{dy}{y}
    {\cal G}_{1\rho}\Big(y,\frac{\hat p}{\kappa}\Big)\,,
    \qquad 
    {{\cal F}}_\rho\Big(x,\frac{\hat p}{\kappa}\Big)
    = 
    - \int_x^1 \frac{dy}{y} \int_y^1 \frac{dy'}{y'}
    {\cal G}_{1\rho}\Big(y',\frac{\hat p}{\kappa}\Big)\,,
    \nonumber
\end{eqnarray}
we find that, surprisingly, the well known Wandzura-Wilczek
relation from deep inelastic $e\,p-$scattering \cite{WW77} have an
operator analogue, namely, the (twist-2 part of) ${\cal
G}_{2\rho}(x,{\hat p}/{\kappa})$ fulfills the same relation as
does the (twist-2 part of the) structure function $g_{2}(x)$ of
DIS,
\begin{eqnarray}
    {\cal G}_{2\rho}\Big(x,\frac{\hat p}{\kappa}\Big)
    \equiv
    {\cal G}^{\;\rm WW}_{2\rho}\Big(x,\frac{\hat p}{\kappa}\Big)
    &=& - \, {\cal G}_{1\rho}\Big(x,\frac{\hat p}{\kappa}\Big)
    + \int_x^1 \frac{dy}{y}
    {\cal G}_{1\rho}\Big(y,\frac{\hat p}{\kappa}\Big)\,.
    \label{g22}
\end{eqnarray}

This means
that we are confronted with a very general structure of the theory
which is independent of taking matrix elements. Instead,
it is a property of the leading twist-2 part of the Fourier transformed
Compton operator.

In addition, another relation holds for the (twist-2 part of the)
operator expression ${\cal G}_{0\rho}(x,{\hat p}/{\kappa})$ which
is unknown in the structure of DIS and is suppressed by a factor
$1/{Q^2}$:
\begin{align}
    {\cal G}_{0\rho}^{\pm}\Big(x,\frac{\hat p}{\kappa}\Big)
    &=
    (x\xi_\pm )\,
    {\cal G}^\pm_{1\rho}\Big(x,\frac{\hat p}{\kappa}\Big)
    - \frac{2x^2+ x\xi_\pm}{[1+x^2 {\hat p}^2/Q^2]^{1/2}}
    \int_x^1\! \frac{dy}{y}
    {\cal G}^\pm_{1\rho}\Big(y,\frac{\hat p}{\kappa}\Big)
    -\frac{2x^2}{[1+x^2 {\hat p}^2/Q^2]^{3/2}}
    \int_x^1\! \frac{dy}{y} \!\int_y^1\! \frac{dy'}{y'}
    {\cal G}^\pm_{1\rho}\Big(y',\frac{\hat p}{\kappa}\Big)\,.
    \label{g02}
\end{align}

 It is remarkable that only one kind of operators, namely
${\cal G}^\pm_{1\rho}(x,{\hat p}/{\kappa})$ or, equivalently,
${\Omega}_{5\rho}^{(2)}(\xi_\pm,{\hat p}/{\kappa})$, determines
the complete structure of the antisymmetric piece of the
(imaginary part of the) Compton operator. Let us emphasize that
this obtains without using any dynamical assumption but, due to
the twist-2 structure of the operator, is a purely geometric
result. Furthermore, these relations respect the complete power
or, after taking matrix elements, target mass corrections which 
result from the (infinite number of) trace terms of the twist-2
Compton operator. 

\section{Symmetric part of the twist-2 Compton operator}
\noindent{\sf A) Entire operator expression} \\
Now we consider the symmetric part of the Compton operator which
is given by \cite{GRE04}
\begin{eqnarray}
{\widehat T}^\twz_{\{\mu\nu\}}\kln{q}
 &=&
 \im\, S^{\mu\nu|\alpha\beta}
 \int \frac{\D^4 \!x}{2\pi^2} \;%
 \frac{\e^{\im qx}\, x_{\alpha}} { {\kln{x^2 - \im\epsilon}^2} }\;%
 \Kln{ O^{\rm tw 2}_{\beta} \Kln{\kappa x, -\kappa x}
 - O^{\rm tw 2}_{\beta} \Kln{- \kappa x, \kappa x} }%
\nonumber \\ %
&=&
 2 \int_0^1 \frac{\d\tau}{\tau} \; \kln{1-\tau + \tau \ln\tau}
 \int \frac{\D^4 \! u}{\kappa^4\tau^4} \;
 {\cal O}_\rho \Big(\frac{u}{\kappa\tau}\Big)
 \, \partial^\rho_u \bigg[\,\frac{2}{[(q + u)^2 + i\epsilon]^3}\times
\nonumber\\
&& \quad
 \bigg\{\!\!\kln{ q^2 \, g_{\mu\nu} - q_\mu q_\nu }  \,
 \left[(u q)^2- u^2q^2 + \hbox{$\frac{1}{2}$} u^2 (q + u)^2 \right]
 + \,  \kln{ u_\mu {q^2}- {q_\mu \kln{uq}} }
       \kln{ u_\nu {q^2}- {q_\nu \kln{uq}} }\bigg\}\;
 \bigg]
 \nonumber\\
&=&
 2\int_{0}^1 d\tau \int_0^1 d\sigma \int_0^1 d\rho
 \int \frac{\D^4 \! u}{\kappa^4} \;
 {\cal O}_\rho \Big(\frac{u}{\kappa}\Big) \;
 \left[q^2\right]^3\, \partial_{\widetilde u}^\rho
 \left\{
 \frac{2 A_{\mu\nu}^{\top}(q,\widetilde u)}{\big[
 \widetilde R(\tau) + \im \epsilon \big]^3}
 + \frac{B_{\mu\nu}^{\top}(q,\widetilde u)}{\big[
 \widetilde R(\tau) + \im \epsilon \big]^2}\!\!
 \right\} ,
\label{Ts1}
\end{eqnarray}
thereby, we introduced the following abbreviations:
\begin{align}
&A_{\mu\nu}^{\top}(q,u)
  = g_{\mu\nu}^{\top}\Big(\frac{qu}{q^2}\Big)^2
  \Big(1 - \frac{q^2 u^2}{(qu)^2}\Big)
  +
  \frac{1}{q^2}u_\mu^{\top} u_\nu^{\top} \,,
\qquad
 B_{\mu\nu}^{\top}(q,u)
  = g_{\mu\nu}^{\top}\frac{u^2}{[q^2]^2}\,,
 \label{abkS1}
\\
& \widetilde R(\tau) = (q + \tau \widetilde u)^2\,,
\qquad
 \widetilde u_\mu = \sigma\rho\, u_\mu\,.
\label{abkS2}
\end{align}

Again, we have to compute the various integrations over $\rho,\,\sigma$ and
$\tau$ partially, together with the derivation w.r.t. $u$, so that finally integrals 
over $1/[R +\im\epsilon]$ only remain. Then, rescaling and
introducing antisymmetric operators ${\cal O}^{(i)}_\alpha\!\kln{u}$
according to Eq.~(\ref{GDA}) the result of the tedious but
straightforward calculation is as follows (cf.~also Ref.~\cite{GRE04}):
\begin{eqnarray}
{\widehat T}^\twz_{\{\mu\nu\}}\kln{q}
 &=&
 \frac{q^2}{2} \int \frac{\D^4 \! u}{\kappa^4} \;
 \Bigg\{
 \frac{q_\alpha}{qu} \Big[
 g_{\mu\nu}^{\top}\; {\cal F}_1^\alpha\kln{\frac{u}{\kappa}}
 +
 \frac{q^2 u_\mu^{\top} u_\nu^{\top}}{(qu)^2-q^2u^2}\;
 {\cal F}_2^\alpha\kln{\frac{u}{\kappa}} \Big]
 \nonumber \\
&& \qquad\qquad
 +
 \Big( \frac{q_\alpha}{qu} - \frac{u_\alpha}{u^2} \Big)
 \Big[
 g_{\mu\nu}^{\top}\;
 {\cal F}_3^\alpha\kln{\frac{u}{\kappa}}
 +
 \frac{q^2 u_\mu^{\top} u_\nu^{\top}}{(qu)^2-q^2u^2}\;
 {\cal F}_4^\alpha\kln{\frac{u}{\kappa}} \Big]
 \nonumber \\
&& \qquad\qquad
 +
 \Big(u_\mu^{\top} g_{\nu\alpha}^{\top}
 +
 u_\nu^{\top} g_{\mu\alpha}^{\top}
 - \,2 \,u_\mu^{\top} u_\nu^{\top} \frac{q_\alpha}{qu} \Big)
 \frac{q^2}{(qu)^2-q^2u^2}\;
 {\cal F}_5^\alpha\kln{\frac{u}{\kappa}}
\Bigg\} \frac{1}{R(1)+ \im\epsilon}
\label{Ts_nonf}
\end{eqnarray}
where $ {\cal F}_{i\alpha}\kln{{u}/{\kappa}}$ are abbreviations of the
following combinations:
\begin{align}
\label{F1}
{\cal F}_{1\alpha}\kln{\frac{u}{\kappa}}
  & =
  {\cal O}^{(0)}_\alpha\!\kln{\frac{u}{\kappa}}
  + \frac{u^2 }{(qu)^2-q^2u^2}\;
  \left(
  (qu+ u^2){\cal O}^{(1)}_\alpha\!\kln{\frac{u}{\kappa}}
  + u^2 \,  {\cal O}^{(2)}_\alpha\!\kln{\frac{u}{\kappa}}
  \right),
\\
\label{F2}
{\cal F}_{2\alpha}\kln{\frac{u}{\kappa}}
  & =
  {\cal O}^{(0)}_\alpha\!\kln{\frac{u}{\kappa}}
  + \frac{3\,u^2 }{(qu)^2-q^2u^2}\;
  \left(
  (qu+ u^2){\cal O}^{(1)}_\alpha\!\kln{\frac{u}{\kappa}}
  + u^2 \,  {\cal O}^{(2)}_\alpha\!\kln{\frac{u}{\kappa}}
  \right),
\\
 \intertext{}
 \label{F3} {\cal F}_{3\alpha}\kln{\frac{u}{\kappa}}
  &=
   -\int_0^1\frac{d\tau}{\tau^2}\,
  \bigg[{\cal F}_{1\alpha}\kln{\frac{u}{\kappa\tau}}
  + \frac{(qu)^2}{(qu)^2-q^2u^2}\;
    {\cal F}_{2\alpha}\kln{\frac{u}{\kappa\tau}}\bigg]
  - \frac{u^2(q+u)^2}{(qu)^2-q^2u^2}
  \int_0^1\frac{d\tau}{\tau^2}\,
  {\cal O}^{(0)}_\alpha\!\kln{\frac{u}{\kappa\tau}}
  \nonumber \\
&  + \frac{2(qu)}{(qu)^2-q^2u^2}
  \int_0^1\frac{d\tau}{\tau^2}
  \left(
  (qu+ u^2)\; {\cal O}^{(0)}_\alpha\!\kln{\frac{u}{\kappa\tau}}
  + u^2\; {\cal O}^{(1)}_\alpha\!\kln{\frac{u}{\kappa\tau}}
  \right)
  \nonumber \\
&  +\frac{u^2}{(qu)^2-q^2u^2}
  \left(
  (q^2 + qu)\; {\cal O}^{(0)}_\alpha\!\kln{\frac{u}{\kappa}}
  +  qu \; {\cal O}^{(1)}_\alpha\!\kln{\frac{u}{\kappa}}
  \right),
\\
\label{F4} {\cal F}_{4\alpha}\kln{\frac{u}{\kappa}}
  &=
  3\, {\cal F}_{3\alpha}\kln{\frac{u}{\kappa}}
  -  \frac{2\,q^2u^2 }{(qu)^2-q^2u^2}\;
  {\cal F}_{5\alpha}\kln{\frac{u}{\kappa}}
 -  \frac{2\,u^2}{(qu)^2-q^2u^2}
  \left(
  (q^2 + qu)\; {\cal O}^{(0)}_\alpha\!\kln{\frac{u}{\kappa}}
  +  qu \; {\cal O}^{(1)}_\alpha\!\kln{\frac{u}{\kappa}}
  \right) ,
\\
{\cal F}_{5\alpha}\kln{\frac{u}{\kappa}}
  &=
  \int_0^1\frac{d\tau}{\tau^2}\;
  {\cal F}_{2\alpha}\kln{\frac{u}{\kappa\tau}}\,.
  \label{F5}
\end{align}
First, we observe the remarkable structural similarity of ${\cal
F}_{1\alpha}$ and ${\cal F}_{2\alpha}$; it is the source of the
Callan-Gross relation which will be shown below. Second, we
observe that ${\cal F}_{3\alpha}$ and ${\cal F}_{4\alpha}$ are
determined by ${\cal F}_{1\alpha}$ and ${\cal F}_{2\alpha}$ and
furthermore contain also those operator combinations of ${\cal
O}^{(0)}_\alpha$ and ${\cal O}^{(1)}_\alpha$ which already
appeared in the trace part, Eq.~(\ref{Trop}). Finally, we
remark that only ${\cal F}_{1\alpha}$ and ${\cal F}_{2\alpha}$
survive when forward matrix elements are taken.

Taking the trace of expression (\ref{Ts_nonf}) one gets
\begin{eqnarray}
\label{TR}
{\widehat T}^\twz_{\rm trace}\kln{q}
 &=&
 \frac{q^2}{2} \int \frac{\D^4 \! u}{\kappa^4} \;
 \bigg\{
 \frac{q_\alpha}{qu}\Big[\,
 3\, {\cal F}_1^\alpha\kln{\frac{u}{\kappa}}
 - {\cal F}_2^\alpha\kln{\frac{u}{\kappa}}\Big]
 \nonumber\\
 &&\qquad\qquad +
 \Big( \frac{q_\alpha}{qu} - \frac{u_\alpha}{u^2} \Big)
 \Big[\,3 \,{\cal F}_3^\alpha\kln{\frac{u}{\kappa}}
 - {\cal F}_4^\alpha\kln{\frac{u}{\kappa}}
 -\frac{2\,q^2u^2}{(qu)^2-q^2u^2} \;
 {\cal F}_5^\alpha\kln{\frac{u}{\kappa}} \Big]
 \bigg\}\frac{1}{R(1)+ \im\epsilon}\,;
\end{eqnarray}
as one convinces oneself, it coincides, modulo $1/R(0)$-terms, with the
former result (\ref{Ftrace0}) which we determined independently by another way.

From this for the symmetric {\em traceless} Compton operator we get
\begin{eqnarray}
\hspace{-.5cm}
{\widehat T}^\twz_{\{\mu\nu\}\rm traceless}\kln{q} &=&
{\widehat T}^\twz_{\{\mu\nu\}}\kln{q}
- \frac{1}{3}\,g_{\mu\nu}^{\top}\; {\widehat T}^\twz_{\rm trace}\kln{q}
\nonumber\\
 &=&
 \frac{q^2}{2} \int \frac{\D^4 \! u}{\kappa^4} \;
 \Bigg\{
 \bigg(
 \frac{1}{3} g_{\mu\nu}^{\top}
  -
 \frac{u_\mu^{\top} u_\nu^{\top}}{(u^{\top})^2}\bigg)
\bigg[\;\frac{q_\alpha}{qu}\Big(
 {\cal F}_2^\alpha\kln{\frac{u}{\kappa}}
 + \frac{(u^{\top})^2}{u^2}
 {\cal F}_4^\alpha\kln{\frac{u}{\kappa}}
 -
 2 {\cal F}_5^\alpha\kln{\frac{u}{\kappa}}\Big)
 - \frac{u_\alpha^\top}{u^2}\,
 {\cal F}_4^\alpha\kln{\frac{u}{\kappa}}\bigg]
 \nonumber \\
&& \qquad\qquad
 +
 \bigg(\frac{2\,g_{\mu\nu}^{\top}u_\alpha^{\top}}{3(u^{\top})^2}
 -
 \frac{u_\mu^{\top} g_{\nu\alpha}^{\top}
 +
 u_\nu^{\top} g_{\mu\alpha}^{\top}}{(u^{\top})^2}\bigg)
  {\cal F}_5^\alpha\kln{\frac{u}{\kappa}}\!
\Bigg\} \frac{1}{R(1)+ \im\epsilon}.
\label{Ts_nonf0}
\end{eqnarray}
Obviously, ${\cal F}_1^\alpha$ and ${\cal F}_3^\alpha$ disappeared
since they are trace parts. Expressions (\ref{Ts_nonf}), (\ref{TR})
and (\ref{Ts_nonf0}) correspond to a kinematic
expansion of related amplitudes. \\

\noindent {\sf B) Restriction to the imaginary part
${\mathrm{Im}}\, {\widehat T}^\twz_{\{\mu\nu\}}\kln{q}$ }
\\
Now, let us use in Eq. (\ref{Ts_nonf}) the representation
(\ref{imR}) of the imaginary part of ${1}/[R(1)+ \im \epsilon]$
and, again, perform the same steps as in Eq.~(\ref{Trop}).
\medskip

\noindent {\it 1. Callan-Gross operator relation: 
${\cal F}_1^\alpha$ and ${\cal F}_2^\alpha$}\\
 First, we consider that contribution to the Compton
operator which remains if forward matrix elements are taken, i.e.,
which is given by the combinations ${\cal F}_1^\alpha$ and ${\cal
F}_2^\alpha$ only; it will be denoted by $^0{\widehat
T}_{\{\mu\nu\}}\kln{q}$. Let us take the imaginary part of
$1/[R(1)+\im\epsilon]$ and let us scale the $u-$variable as
previously has been done,
\begin{align}
\hspace{-.25cm}
{\mathrm{Im}}\, ^0{\widehat T}^\twz_{\{\mu\nu\}}\kln{q}
    =&~
    {\mathrm{Im}}\, ^0{\widehat T}^+_{\{\mu\nu\}}\kln{q}
    +
    {\mathrm{Im}}\, ^0{\widehat T}^-_{\{\mu\nu\}}\kln{q},
    \nonumber\\
{\mathrm{Im}}\, ^0{\widehat T}^\pm_{\{\mu\nu\}}\kln{q}
    =&~
    \frac{-\pi}{2}\int_{{\cal H}^3} \frac{\D^4 \hat p}{\kappa^4}
    \,\delta({\hat p}^2 - 1)\,
    \int \d t\,t^2\;\delta\big(t - \tilde\xi_\pm\big)
    \frac{q^\alpha}{q\hat p}
 \nonumber\\
 \hspace{-.25cm}
 &\;\times\;
 \bigg[
 \frac{q^2 g_{\mu\nu}^{\top} }{[(q{\hat p})^2 - q^2 {\hat p}^2]^{1/2}}
 \bigg\{
 {\cal O}^{(0)}_\alpha\!\kln{\frac{t{\hat p}}{\kappa}}
 \pm \frac{t\,{\hat p}^2}{[(q{\hat p})^2 - q^2 {\hat p}^2]^{1/2}}
 {\cal O}^{(1)}_\alpha\!\kln{\frac{t{\hat p}}{\kappa}}
  + \frac{t^2\,({\hat p}^2)^2}{[(q{\hat p})^2-q^2{\hat p}^2]}
 {\cal O}^{(2)}_\alpha\!\kln{\frac{t{\hat p}}{\kappa}}
 \bigg\}
 \nonumber\\
\hspace{-.25cm}
 &\;+ \;
 \frac{[q^2]^2\,{\hat p}_\mu^{\top} {\hat p}_\nu^{\top}}
 {[(q{\hat p})^2 - q^2 {\hat p}^2]^{3/2}}
 \bigg\{
  {\cal O}^{(0)}_\alpha\!\kln{\frac{t{\hat p}}{\kappa}}
  \pm \frac{3\,t\,{\hat p}^2}{[(q{\hat p})^2 - q^2 {\hat p}^2]^{1/2}}
  {\cal O}^{(1)}_\alpha\!\kln{\frac{t{\hat p}}{\kappa}}
  + \frac{3\,t^2\,({\hat p}^2)^2}{[(q{\hat p})^2-q^2{\hat p}^2]}
  {\cal O}^{(2)}_\alpha\!\kln{\frac{t{\hat p}}{\kappa}}
 \bigg\}
 \bigg]\phantom{\Bigg|}
 \nonumber\\
\intertext{}
 \hspace{-.25cm}
 = &~
  \frac{\pi}{2}\int_{{\cal H}^3} \frac{\D^4 \hat p}{\kappa^4}
  \,\delta({\hat p}^2 - 1)\frac{q^\alpha}{q\hat p}
 \label{Ts_fim}\\
 \hspace{-.25cm}
 &\times
  \bigg[
 \frac{x g_{\mu\nu}^{\top}}{[1+x^2 {\hat p}^2/Q^2]^{1/2}}
 \bigg\{
  {\Omega}_\alpha^{(0)}\Big(\xi_\pm,\frac{\hat p}{\kappa}\Big)
  \pm \frac{x{\hat p}^2/Q^2}{[1+x^2 {\hat p}^2/Q^2]^{1/2}}
  {\Omega}_\alpha^{(1)}\Big(\xi_\pm,\frac{\hat p}{\kappa}\Big)
  + \frac{x^2 ({\hat p}^2/Q^2)^2}{[1+x^2 {\hat p}^2/Q^2]}
  {\Omega}_\alpha^{(2)}\Big(\xi_\pm,\frac{\hat p}{\kappa}\Big)
 \bigg\}
 \nonumber\\
 \hspace{-.25cm}
 &\; -
 \frac{x^3 {\hat p}_\mu^{\top}
 {\hat p}_\nu^{\top}/Q^2}{[1+x^2 {\hat p}^2/Q^2]^{3/2}}
 \bigg\{
  {\Omega}_\alpha^{(0)}\Big(\xi_\pm,\frac{\hat p}{\kappa}\Big)
  \pm \frac{ 3\, x {\hat p}^2/{Q^2}}{[1+x^2 {\hat p}^2/Q^2]^{1/2}}
  {\Omega}_\alpha^{(1)}\Big(\xi_\pm,\frac{\hat p}{\kappa}\Big)
  + \frac{3\, x^2 ({\hat p}^2/Q^2)^2}{[1+x^2 {\hat p}^2/Q^2]}
  {\Omega}_\alpha^{(2)}\Big(\xi_\pm,\frac{\hat p}{\kappa}\Big)
 \bigg\}\bigg]
 \nonumber
 \end{align}
where we used the distribution amplitudes (\ref{Phi_tr})
and extended them to arbitrary order $i$:
\begin{eqnarray}
 {\Omega}_\alpha^{(0)}(t,\hat p)
 \equiv
 t^2\,{\cal O}^{(0)}_\alpha\!\kln{t\,{\hat p}}\,,
 \qquad
 {\Omega}_a^{(i)}(t,\hat p) \equiv
 \int_t^1 dy\,{\Omega}_\alpha^{(i-1)}(y,\hat p)
 \quad {\rm for} \quad i \geq 1 \,.
 \nonumber
\end{eqnarray}

Now, let us look for possible relations between the kinematical
structures of expression (\ref{Ts_fim}).

 Formally, the
structure in the angular bracket of expression (\ref{Ts_fim}) is
completely analogous to the corresponding one of the structure
functions $W_1$ and $W_2$ in DIS as given by Georgi and Politzer
\cite{GP76}. Therefore, let us introduce the operators
${\cal W}_1$ and ${\cal W}_2$ according to
\begin{eqnarray}
 \label{TS}
 {\rm Im}\,^0\widehat{\cal T}^\twz_{\{\mu\nu\}}\kln{q}
 &=&
 \pi\int_{{\cal H}^3} \frac{\D^4 \hat p}{\kappa^4}
  \,\delta({\hat p}^2 - 1)\frac{q^\alpha}{q\hat p}
 \bigg[- g_{\mu\nu}^{\top}\,
 {\cal W}_{1\alpha} \Big(x,\frac{\hat p}{\kappa}\Big)
 + \frac{{\hat p}_\mu^{\top} {\hat p}_\nu^{\top}}{{\hat p}^2} \,
 {\cal W}_{2\alpha}\Big(x,\frac{\hat p}{\kappa}\Big)
 \bigg]\,.
\end{eqnarray}
Then, we observe that both operators may be written as:
\begin{eqnarray}
{\cal W}_{i\alpha}\Big(x,\frac{\hat p}{\kappa}\Big)
 &=&
  {\cal W}_{i\alpha}\Big(\xi_+,x;\frac{\hat p}{\kappa}\Big)
 +{\cal W}_{i\alpha}\Big(\xi_-,x;\frac{\hat p}{\kappa}\Big)
 \qquad {\rm for} \qquad i = 1,2,L\,,
\nonumber\\
 x\,
 {\cal W}_{1\alpha}\Big(\xi_\pm,x;\frac{\hat p}{\kappa}\Big)
 &=&
 \frac{(q\hat p)}{{\hat p}^2}\,
 {\cal W}_{2\alpha}\Big(\xi_\pm,x;\frac{\hat p}{\kappa}\Big)
 +
 x^2\,\frac{{\hat p}^2}{Q^2}
 \bigg[\bigg(
 x \frac{\pd}{\pd x}
 + x^2
 \frac{\pd^2}{\pd x^2}\;\frac{x}{\xi_\pm} \bigg)
 \bigg(
 \frac{x}{\xi_\pm}
 \frac{{\Omega}_\alpha^{(2)}\big(\xi_\pm,{\hat p}/{\kappa}\big)}
 {\sqrt{1+x^2 {\hat p}^2/Q^2}}
 \bigg)
 \bigg]
\label{for1}\\
&=&
 (1+x^2 {\hat p}^2/Q^2) \,
 \frac{(q\hat p)}{{\hat p}^2}\,
 {\cal W}_{2\alpha}\Big(\xi_\pm,x;\frac{\hat p}{\kappa}\Big)
 - x\,
 {\cal W}_{L\alpha}\Big(\xi_\pm,x;\frac{\hat p}{\kappa}\Big)\,,
 \nonumber
\\
 \frac{(q\hat p)}{{\hat p}^2}\,
 {\cal W}_{2\alpha}\Big(\xi_\pm,x;\frac{\hat p}{\kappa}\Big)
 &=& x^2 \frac{\pd^2}{\pd x^2}
 \bigg(
 \frac{x^2}{\xi^2_\pm}
 \frac{{\Omega}_\alpha^{(2)}\big(\xi_\pm,{\hat p}/{\kappa}\big)}
 {\sqrt{1+x^2 {\hat p}^2/Q^2}}\bigg)\,,
\label{for2}\\
 x\,
 {\cal W}_{L\alpha}\Big(\xi_\pm,x;\frac{\hat p}{\kappa}\Big)
 &=&
 - x^2\,\frac{{\hat p}^2}{Q^2}\;x \frac{\pd}{\pd x}
 \bigg(\frac{x}{\xi_\pm}
 \frac{{\Omega}_\alpha^{(2)}\big(\xi_\pm,{\hat p}/{\kappa}\big) }
 {\sqrt{1+x^2 {\hat p}^2/Q^2}}\bigg)\,.
 \label{forL}
\end{eqnarray}
Obviously, equation (\ref{for1}) is the operator version of the
well-known (target mass corrected) {\sf Callan-Gross relation}
\cite{CG69}.
\medskip

\noindent {\it 2. Remaining part:  ${\cal F}_3^\alpha$, 
${\cal F}_4^\alpha$ and ${\cal F}_5^\alpha$ }\\
Now, let us consider the remaining part $^1{\widehat
T}_{\{\mu\nu\}}\kln{q}$ of the symmetric Compton operator which
will vanish if forward matrix elements are taken. However, since
the combinations
  are given in terms of ${\cal F}_1^\alpha$, ${\cal F}_2^\alpha$
  and the combinations
 ${\cal V}_{0\alpha}$ and ${\cal V}_{1\alpha}$,
modulo a term which vanishes according to the relations (\ref{+}),
we have not to introduce new structure functions. For that part of
the Compton operator, taking into account the relation
$[{1+x^2{\hat p}^2/Q^2}]/({x^2/Q^2})\equiv ({\hat
p}^\mathrm{T})^2$, we obtain:
\begin{align}
{\mathrm{Im}}\, ^1{\widehat T}^\twz_{\{\mu\nu\}}\kln{q}
  &=\;
  {\mathrm{Im}}\, ^1{\widehat T}^+_{\{\mu\nu\}}\kln{q}
  +
  {\mathrm{Im}}\, ^1{\widehat T}^-_{\{\mu\nu\}}\kln{q}\,,
  \nonumber\\
 {\rm Im}\,^1{\widehat T}^\pm_{\{\mu\nu\}}\kln{q}
 &=\,
 {\pi}\int_{{\cal H}^3} \frac{\D^4 \hat p}{\kappa^4}
  \,\delta({\hat p}^2 - 1)
 \Bigg\{2\, \Big( \frac{q_\alpha}{q{\hat p}}
     - \frac{{\hat p}_\alpha}{{\hat p}^2} \Big)\;
  \frac{{\hat p}_\mu^{\top}{\hat p}_\nu^{\top}}{({\hat p}^\mathrm{T})^2}\;
 x\,\frac{{\hat p}^2}{Q^2}\,
 {\cal V}_{1}^{\alpha}
 \Big(\xi_\pm,x;\frac{\hat p}{\kappa}\Big)
 \nonumber \\
& \qquad\qquad
 +
 \frac{1}{{\hat p}^2}\bigg(
 {\hat p}_\mu^{\top} g_{\nu\alpha}^{\top}
 +{\hat p}_\nu^{\top}g_{\mu\alpha}^{\top}
 - 2\,
 \frac{{\hat p}_\mu^{\top} {\hat p}_\nu^{\top}}{({\hat p}_\mu^{\top} )^2}
 {\hat p}_\alpha^{\top}
 \bigg)
 \int_0^1\frac{d\tau}{\tau^2}\,
  {\cal W}_{2}^\alpha
  \Big(\frac{\xi_\pm}{\tau},x;\frac{\hat p}{\kappa}\Big)
\nonumber\\
&  \qquad\qquad 
 + \Big( \frac{q_\alpha}{q{\hat p}}
     - \frac{{\hat p}_\alpha}{{\hat p}^2} \Big)
 \bigg(
  g_{\mu\nu}^{\top}
  -
  3\frac{{\hat p}_\mu^{\top}{\hat p}_\nu^{\top}}{({\hat p}^\mathrm{T})^2}\;
  \bigg) 
  \bigg[
  \int_0^1\frac{d\tau}{\tau^2}
  \bigg(
   {\cal W}_{1}^\alpha
   \Big(\frac{\xi_\pm}{\tau},x;\frac{\hat p}{\kappa}\Big)
   + \frac{1}{x}\frac{q\hat p}{{\hat p}^2}\,
   {\cal W}_{2}^\alpha
   \Big(\frac{\xi_\pm}{\tau},x;\frac{\hat p}{\kappa}\Big)
   \bigg)
  \nonumber \\
&  \qquad\qquad \qquad\quad
~~+ 2 \int_0^1\frac{d\tau}{\tau^2}
  \bigg(
  {\cal V}_{0}^{\alpha}
  \Big(\frac{\xi_\pm}{\tau},x;\frac{\hat p}{\kappa}\Big)
  +x\,\frac{{\hat p}^2}{Q^2}\,
  {\cal V}_{1}^{\alpha}
  \Big(\frac{\xi_\pm}{\tau},x;\frac{\hat p}{\kappa}\Big)
  \bigg)
+ x\,\frac{{\hat p}^2}{Q^2}\,
  {\cal V}_{1}^{\alpha}\Big(\xi_\pm,x;\frac{\hat p}{\kappa}\Big)
  \bigg]\Bigg\}
\,. \label{TSext}
\end{align}
Despite being quite complicated, this piece of the imaginary part
of the symmetric Compton operator -- which does not contribute if
forward matrix elements are taken -- is entirely given by ${\cal
W}_{i}^\alpha({\xi_\pm},x;{\hat p}/{\kappa})$ and ${\cal
V}_{i}^\alpha({\xi_\pm},x;{\hat p}/{\kappa})$.
\bigskip

\noindent{\sf C) Operator relations between ${\cal V}_{i}^\alpha$
and ${\cal W}_{i}^\alpha$}
\\
Up to now we found the operator analogue (\ref{for1}) of the 
Callan-Gross relation.
 Now let us show that ${\cal V}_{0}^\alpha$ and
 ${\cal V}_{1}^\alpha$ can be expressed by two of the structure
 functions ${\cal W}_{i}^\alpha$. Namely, taking the trace
 of Eq.~(\ref{TS}) and comparing it with the expression (\ref{FTraceX})
 one finds
 \begin{align}
 2 \mathcal{V}_{0}^\alpha({\xi_\pm},x;{\hat p})
 =
 \mathcal{W}_{L}^\alpha({\xi_\pm},x;{\hat p})
 -\,2 \mathcal{W}_{1}^\alpha({\xi_\pm},x;{\hat p})\,.
 \label{v0}
 \end{align}
Furthermore, taking the trace of expression (\ref{TSext}) one
finds $2\pi\,x\,({\hat p}^2/{Q^2})\, {\cal V}_{1}^{\alpha}
 (\xi_\pm,x;{\hat p}/{\kappa})$, i.e., the remaining part
  of (\ref{FTraceX}), as it should be.
Now, according to the definition (\ref{H1}), making use of
Eqs.~(\ref{rel1}) -- (\ref{rel4}) and of expression (\ref{forL}),
we get
\begin{eqnarray}
 {\mathcal{V}}_{1\alpha}(\xi_\pm,x;{\hat p})
 &=&
 x\frac{\pd}{\pd x}
 \frac{x \Omega^{(1)}_\alpha(\xi_\pm,{\hat p})}{\sqrt{1+x^2{{\hat p}}^2/Q^2}}
 =
 \mp\, x\frac{\pd}{\pd x} \left[
 \frac{x}{\xi_\pm} x\frac{\pd}{\pd x}
 \Omega^{(2)}_\alpha(\xi_\pm,{\hat p})
 \right]
 \nonumber\\
 &=&
 \pm\, x\frac{\pd}{\pd x} \frac{Q^2}{{\hat p}^2}\left[
 \frac{\sqrt{1+x^2{\hat p}^2/Q^2}}{x}\,
 {\mathcal{W}}_{L\alpha}(\xi_\pm,x;{\hat p})
 -\frac{x \xi_\pm \,{\hat p}^2/Q^2}{\sqrt{1+x^2{\hat p}^2/Q^2}}
 \int_x \frac{\d y}{y^2}\,
 {\mathcal{W}}_{L\alpha}(\xi_\pm(y),y;{\hat p})
 \right]\,,
\end{eqnarray}
so that, observing finally the equality
$\pm \sqrt{1+x^2{\hat p}^2/Q^2} = 1 + x \xi_\pm {\hat p}^2/Q^2$ again,
we obtain
 \begin{eqnarray}
 {x} \frac{{\hat p}^2}{Q^2}
 {\mathcal{V}}_{1\alpha}(\xi_\pm,x;{\hat p})
 &=&
 [1 + x \xi_\pm {\hat p}^2/Q^2]\,\left[\;
 x\frac{\pd}{\pd x}\,
 {\mathcal{W}}_{L\alpha}(\xi_\pm,x;{\hat p})
 -
 \frac{1 -x \xi_\pm{\hat p}^2/Q^2}{1+x^2{\hat p}^2/Q^2}
 {\mathcal{W}}_{L\alpha}(\xi_\pm,x;{\hat p})
 \right.
 \nonumber\\
 &&\qquad\qquad\qquad\qquad \left.
 - \;
 \frac{x^3 {\hat p}^2/Q^2}{[1+x^2{\hat p}^2/Q^2]^2}
 \int_x \frac{\d y}{y^2}\,
 {\mathcal{W}}_{L\alpha}(\xi_\pm(y),y;{\hat p})
 \right].
 \label{v1}
 \end{eqnarray}

With these results we find that the structure of the imaginary part 
of the symmetric twist-2 Compton operator is determined completely by
only two operator valued structure functions, namely
 ${\mathcal{W}}_{1\alpha}(\xi_\pm,x;{\hat p})$ and
 ${\mathcal{W}}_{L\alpha}(\xi_\pm,x;{\hat p})$.

\section{Conclusions}

The investigation of quantities of definite geometric twist is
continued by consideration of the `complete' twist-2 Compton
operator, i.e. given off-cone and taking into account subtraction
of all trace terms. We have shown that it has a hidden structure
which allows to derive {\em without further approximations} and
{\em without relation to higher (geometric) twist} Compton operators 
an operator analog of the Wandzura-Wilczek relation (\ref{g22}) and 
of the (target mass corrected) Callan-Gross relation (\ref{for1}); 
additional structural relations (\ref{g02}), (\ref{v0}) and (\ref{v1}), 
up to now unobserved, occur as well. According to their derivation one is 
led to conclude that our version of the Wandzura-Wilczek and Callan-Gross
relations -- also in the case non-forward scattering -- are
determined completely geometrically, i.e., solely by the twist
structure of the operator behind the scattering amplitudes.

All these relations -- together 
with the operator expressions (\ref{Ftrace0}), (\ref{Tas501}) and 
(\ref{Ts_nonf}) of the trace, antisymmetric and symmetric part of the 
entire twist-2 Compton operator -- contain the {\em complete power 
corrections} resp. {\em target mass corrections} of the twist-2 Compton 
operator. The proof of all these relations goes,
with more or less obvious changes, along the same lines as has
been done for the non-forward virtual Compton amplitude in
Ref.~\cite{GRE04}. Taking matrix elements of the Compton operator
these structural relations remain intact and lead to the
corresponding relations for the generalized structure functions
and distribution amplitudes. Obviously, they hold equally well for
the forward \cite{GL01} and non-forward scattering amplitudes
\cite{BR00}, but also more complicated matrix elements, e.g., when
a meson occurs in the final state \cite{BEGR02} or when
diffractive scattering \cite{BR01} is considered.

A remarkable result we found was that only the {\em three independent operator 
valued structure functions}, namely 
 ${\mathcal{G}}_{1\alpha}(\xi_\pm,x;{\hat p})$,
 ${\mathcal{W}}_{1\alpha}(\xi_\pm,x;{\hat p})$ and
 ${\mathcal{W}}_{L\alpha}(\xi_\pm,x;{\hat p})$, completely determine 
 the structure of the imaginary part of the twist-2 Compton operator. 
In the course of that calculations it was essential to introduce
analogs of Bjorken and Nachtmann variables $x$ and $\xi$,
respectively, together with a scaling variable $t$. This latter
variable measures, when matrix elements are taken, the collinear
momentum of the corresponding scattering process, i.e., that part
of the momentum pointing in the direction ${\cal P}\equiv \sum_i P_i$.
Furthermore, let us mention, that $x \rightarrow Q^2/q\cal P$ and
that the essential $Q^2-$dependence appears only through ${\hat
p}^2 / Q^2 \rightarrow {\cal P}^2 / Q^2$, where the arrow shows
the corresponding dependence of the matrix elements.

Concerning phenomenological applications our expressions for the entire 
Compton operator contain  mass corrections for the standard deep inelastic 
scattering and reproduce the existing results. They are also applicable 
to many different exclusive processes like DVCS and, more general,
$ e^- + N = e^- + N +$ particle~1 + particle~2 + ... + particle~$n$ as well as
$ e^+ + e^- = $ particle~1 + particle~2 + ... + particle~$n$, thereby
using the kinematics of generalized Bjorken region and forming
corresponding matrix elements of the Compton operator. Of course
at present current conservation remains an open problem which restricts
that applicability. It seems to be likely that this can be resolved, at
least approximate, by including a center coordinate into the definition of 
the Compton operator and, as a consequence, adding kinematical twist-3 
contributions analogous to the procedures in Refs.~\cite{BM00,APT00,RW00,BB95} 
or by performing a `quark spin rotation' as proposed in \cite{RW01} or using 
`rotational invariance' as proposed in \cite{AT01}. However, this has not been 
considered here. Especially, because of the complexity of the (geometric)
twist-3 Compton operator the study of its structural relations has to be postponed.

\bigskip
\medskip

\noindent
{\large {\bf Acknowledgement}}\\
The authors are indebted to J.~Bl\"umlein for various useful discussions.
They also thank an anonymous referee for some questions which led them to
a more detailed and sharpened formulation.
D.~Robaschik thanks Institute of Theoretical Physics
of Leipzig University for kind hospitality and Graduate College
"Quantum field theory" for financial support.


\end{document}